\documentclass[12pt, a4paper]{article}
\usepackage[english]{babel}
\usepackage[utf8]{inputenc}
\usepackage[autostyle, english = american]{csquotes}
\usepackage[T1]{fontenc}
\usepackage{times}
\usepackage{amsmath,amssymb,amsfonts}
\usepackage{graphicx}
\usepackage{subfig}
\usepackage[dvipnames]{xcolor}
\usepackage{caption}
\usepackage[margin=2cm]{geometry}
\usepackage[doublespacing]{setspace}
\usepackage{textgreek}
\usepackage[utf8]{inputenc}
\usepackage{authblk}
\usepackage{cleveref}
\usepackage[bottom]{footmisc}
\title{\vspace{-15mm} Regular charged black hole in massive gravity as heat engine }
\author[1]{Shubham Sharma \footnote{shubham2019gb@gmail.com}}\author[1]{Amruta Desai \footnote{amrutadesai011@gmail.com}} \affil[1]{Department of  Physics, National Institute of Technology, Srinagar, Jammu \& Kashmir}
\author[1]{Prince A. Ganai \footnote{princeganai@nitsri.net}}
\date{}
\begin{document}
\maketitle
\begin{abstract}
 We examine the effect of addition of graviton mass on the efficiency of a  regular charged black hole (RCB). In doing so, we make few comments on the critical behaviour of the system and calculate the relevant thermodynamic quantities such as entropy, Hawking's temperature and heat capacity. We confirm that graviton mass has a positive effect on the efficacy of such a heat engine under consideration.      
\end{abstract} 
\pagebreak
\section{Introduction}
Classically, black holes are simple objects parametrized with some basic quantities such as mass, angular momentum and charge \cite{1}. In this state they are known to behave like perfect absorbers that would suck every matter falling into it without emitting back anything. However a whole new branch of physics emerges once quantum effects are taken into account and black holes become one of the most important links connecting gravitation, thermodynamics and quantum theory. This was first realized through the early works of Hawking and Bekenstein \cite{2,3,4}, wherein the concepts of temperature and entropy were introduced for the first time under the light of black hole physics. Subsequently, the four laws describing thermodynamics of a black hole were formulated \cite{5} \footnote{A close connetion of the laws of thermodynamics with that of gravity has been discussed in \cite{6,7}}. This was then followed by the discovery of existence of the canonical ensemble for AdS geometry asymptotically \cite{8}, which means that the black hole [BH] can exhibit critical region where it undergoes phase transitions. Note that this is not the case for asymptotically flat space geometry. This realization opened door to investigations into wide variety of fields such as the \emph{AdS/CFT} duality \cite{9,10,11}, quantum chromodynamics \cite{12}, condensed matter physics \cite{13,14}, and geometrical approaches to BH thermodynamics \cite{15}. Along with this, there has been addition of important concepts of \emph{pressure} and \emph{volume} to the existing BH parameters which subsequently led to some significant changes in the laws for BH thermodynamics, especially from quantum physics point of view.     
\subsection{Expansion of BH thermodynamics} The original first law of BH thermodynamics, that exploits the quanutm relation between temperature \emph{T} and surface gravity  $\kappa$ and that of entropy \emph{S} and area of the black hole, for a charged rotating BH has the following form :
\begin{equation} \label{eqn : 1}
    \delta M = T\delta S + \Omega\delta J + \phi \delta Q \ ,
\end{equation} 
where \emph{M} is the mass identified with the internal energy \emph{U}, $\Omega$ and \emph{J} is angular velocity and angular momentum associated with rotating black hole and $\phi$ and \emph{Q} is the electric potential and charge associated with the charged BH. But equation \ref{eqn : 1} does not yet include the \emph{pressure} and \emph{volume} terms which are commonly found in thermodynamics. It was by then well established that the cosmological constant $\Lambda$ can be treated as a thermodynamical BH parameter \cite{16,17}. Consequently, the \emph{pressure} of the BH was interpreted in terms of a negative cosmological constant and the corresponding $P \delta V$ term was finally added to the original first law  \cite{18}, leading to  a more generalized first law of the following form :
\begin{equation}
    \delta M = T\delta S + V \delta P + \Omega\delta J + \phi \delta Q\ ,
\end{equation} where \begin{equation*}
 P = \frac{-\Lambda}{8\pi}
\end{equation*} and the corresponding conjugate variable volume is \cite{19,20},
\begin{equation*}
V \equiv \left( \frac{\partial M}{\partial P}\right)_{S,Q,J}
\end{equation*} This also changed the interpretation of mass \emph{M} to the enthalpy \emph{H} \cite{21} of the BH, which now involves both the internal energy \emph{U} and energy due to the \emph{PV} term. Expansion of BH thermodynamics in such a way leads to a new extended picture of the phase space that allows us to study phase transitions with a renewed perspective. Such an investigation has led to an enhanced identification between the AdS black holes and van der Waals liquid-gas system \cite{22,23} and other interesting discoveries \cite{24,25,26,27,28}.
\subsection{Holographic heat engines}
With this kind of framework at hand, Johnson put forth the idea of "holographic heat engines" \cite{29}, wherein he utilises the \emph{pdV} term to obtain mechanical work from heat energy, black holes being the working substance. For such a thermodynamic system, work can be defined in terms of a cycle in the \emph{P - V} plane with \emph{$Q_H$} and \emph{$Q_C$} being the net input and output heat respectively and the net mechanical work is written as \emph{W} =   \emph{$Q_H$} - \emph{$Q_C$}. The heat engine's efficiency is defined in the usual thermodynamical sense as $\eta = W/Q_H = 1 - Q_C / Q_H$. As we know, the efficiency of Carnot's engine corresponds with the maximum achievable efficiency given by,
\begin{equation*}
    \eta_c = 1 - \frac{Q_C}{Q_H} = 1 - \frac{T_C}{T_H}
\end{equation*} where $T_C$ and $T_H$ are lower and higher temperature of the reservoir respectively. \par In this paper, we develop a more generalised version of thermodynamic heat engine by incorporating the idea of massive gravity and study the effects of such a modification on the efficiency of the heat engine. There are several arguments one can make in support of such a generalisation : The theory of general relativity (GR) is known to describe the gravitational force with a good agreement with the observations. However the search for alternative theories has been an on-going quest since the formulation of GR. From a modern particle physics
perspective, GR is a unique theory for a massless spin-2 particle. Taking GR as starting point, alternative theories intend to modify or extend GR to different scales in a way as to account for the observational anomalies. One such modification maintains the notion of gravity being propagated by a spin-2 particle while giving a mass to this particle, that is the so called "Massive gravity". The first attempt to add a mass to the graviton was made by Fierz and Pauli \cite{30}, who in their work showed the presence of an extra degree of freedom (called ghost) in a generic Lorentz-invariant modification of gravity. Unfortunately, presence of an unwanted mode was discovered, named as the vDVZ discontinuity \cite{31,32} that represented failure of a massive gravity theory to reduce itself into general relativity (a massless spin-2 field theory with two propagating degree of freedom ) in the massless limit. This pathology naturally posed a threat to one of the most important underlying principle when it comes to deforming a theory : Continuity in physical predictions of a theory in the parameters of the theory. This issue was then resolved by Vainshtein in \cite{33} by proposing a mechanism through which nonlinearities managed to hid the extra degrees of freedom, at short distances and has been successfully applied to several models \cite{34,35,36}.
\par Thus, to summarize, the cosmological models motivated by GR are plagued by issues like the
cosmological constant problem, late time acceleration of universe, issues at the interface of gravity and quantum mechanics, etc., forcing us to look for modifications such as mentioned above. Massive gravity theories are of immense interest as an alternative to GR, strongly motivated on both theoretical and phenomenological grounds \cite{37,38,39}.
 Undoubtedly, its inclusion in the study of the efficiency of BH heat engines would result into some interesting insights. Also there are enough evidences of such impact of a massive graviton, for example the presence of van der Waals phase transitions for non-spherical geometries, in holography \cite{40}, in the resolution to the cosmological constant problem \cite{41,42}, for enhanced polarization of gravitational waves \cite{43} and many more. It would be interesting to find out the effect of such highly influential graviton mass on the efficacy of the heat engines. We consider a regular charged black hole (RCB) solution for such an investigationand and for additional work along this direction for different black hole systems see\cite{44,45,46}. In doing so, we also calculate temperature, pressure and specific heat of the RCB heat engine. \par The paper is organised as follows : In the next section, we begin by constructing a RCB heat engine in massive gravity in extended phase space and draw some important results through their plots. In section \ref{section :3} we turn to our main aim that is to study the effect of inclusion of graviton mass on the efficiency of the RCB heat engine. In the last section, we summarize our results.
\section{Regular charged black holes in massive gravity as heat engines}
We consider the following metric, representing the 4-dimensional spherically symmetric RCB in massive gravity,
\begin{equation}
    ds^2 = -f(r)dt^2 + f(r)^{-1}(r)dr^2 + r^2 (d\theta^2 + sin^2\theta d\phi^2)
\end{equation} where the metric function \emph{f(r)} consists of terms due to electric charge and massive gravity \cite{47},
\begin{equation} \label{eqn : 4}
    f(r) = 1 - \frac{2M}{r}e^{\left(\frac{-q^2}{2Mr}\right)} - \Lambda\frac{r^2}{3} + m^2c_1 + 2m^2c_2
\end{equation} Here \emph{M} is related to the total mass of the black hole, \emph{q} is the electric charge and \emph{m} is the mass of the graviton. The Hawking's temperature of the black hole is obtained using equation \ref{eqn : 4} and the relation,
\begin{equation}
T_H = \frac{1}{4\pi}\left[\frac{2M e^{\left(\frac{-q^2}{2Mr}\right)} }{r^2} - \frac{q^2 e^{\left(\frac{-q^2}{2Mr}\right)}}{r^3} - \frac{2\Lambda r}{3} + c_1m^2\right]
\end{equation}
\begin{figure}
    \centering
    \includegraphics[width=10cm,height=7cm]{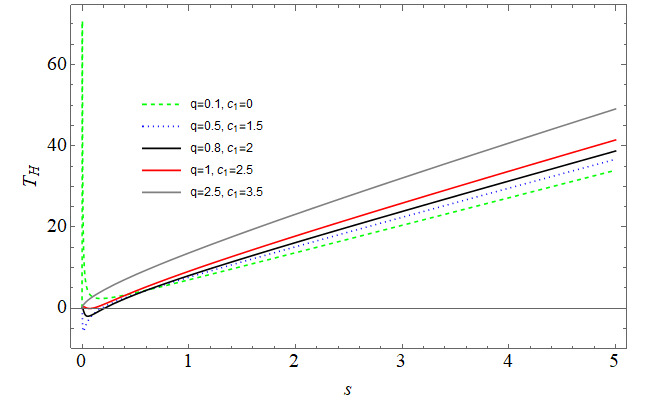}
    \caption{Temperature of RGB in terms of entropy \emph{S}}
    \label{fig : 1}
\end{figure} \\ We can write the temperature in terms of entropy \emph{S} and pressure \emph{P} of the BH,
\begin{equation} \label{eqn : 6}
T_H = \frac{1}{4\sqrt{\pi S}} \left[ \ \frac{2M \sqrt{\pi}}{\sqrt{S}}e^{\left(\frac{-q^2 \sqrt{\pi}}{2M \sqrt{S}}\right)} - \frac{q^2 \pi}{S}e^{\left(\frac{-q^2 \sqrt{\pi}}{2M \sqrt{S}}\right)} + \frac{16 SP}{3\pi} + c^2m\sqrt{\frac{S}{\pi}}\right ]    
\end{equation} 
We then obtain the pressure \emph{P} by using equation \ref{eqn : 6} in terms of \emph{$T_H$} and \emph{S} as,
\begin{equation} \label{eq : 7}
P = \frac{3\pi}{16 S}\left[T_H 4\sqrt{\pi S} - 2M \sqrt{\pi} \ e^{\left(\frac{-q^2 \sqrt{\pi}}{2M \sqrt{S}}\right)} + \frac{q^2 \pi}{S}e^{\left(\frac{-q^2 \sqrt{\pi}}{2M \sqrt{S}}\right)} - c^2m\sqrt{\frac{S}{\pi}}\right]    
\end{equation} We can reformulate equation \ref{eq : 7} for \emph{P} in terms of volume \emph{V} as well,
\begin{equation}
\begin{split}
P &= \frac{9V}{64\pi}^{-2/3}\biggl[ T_H4\pi\left(\frac{3V}{4\pi}\right)^{1/3} - 2M\sqrt{\pi}\exp{\left(\frac{-q^2 \left(\frac{3V}{4\pi}\right)^{-1/3}}{2M}\right)} + \frac{q^2 \pi \exp{\left(\frac{-q^2 \left(\frac{3V}{4\pi}\right)^{-1/3}}{2M}\right)}}{\pi\left(\frac{3V}{4\pi}\right)^{2/3}} \\ &- c^2m\left(\frac{3V}{4\pi}\right)^{1/3}\biggr]
\end{split}
\end{equation} 
\begin{figure}
    \centering
    \includegraphics[width=10cm,height=7cm]{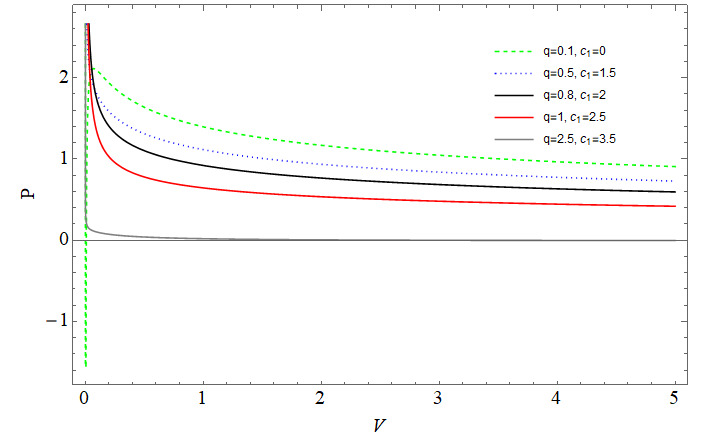}
    \caption{\emph{P-V} diagram demonstrating the heat engine cycle. Here we have taken \emph{m}  = 0.2.}
    \label{fig : 2}
\end{figure}Figure \ref{fig : 2} above shows \emph{P-V} graph using equation \ref{eq : 7} that highlights the rectangular heat engine cycle with various isotherms and varying values of parameters \emph{q} and \emph{c}. To study the critical behaviour of the black hole, we simply obtain the expression for specific heat and plot it against entropy \emph{S} and observe the corresponding phase transitions of the system. So next we obtain the heat capacity at constant pressure by using the standard thermodynamic relation,
\begin{equation}
\begin{split}
C_P &= T_H \left.\left( \frac{\partial S}{\partial T_H}\right) \right\vert_P  \\ &=\frac{1}{4\sqrt{\pi S}} \left[ \ \frac{2M \sqrt{\pi}}{\sqrt{S}}e^{\left(\frac{-q^2 \sqrt{\pi}}{2M \sqrt{S}}\right)} - \frac{q^2 \pi}{S}e^{\left(\frac{-q^2 \sqrt{\pi}}{2M \sqrt{S}}\right)} + \frac{16 SP}{3\pi} + c^2m\sqrt{\frac{S}{\pi}}\right]\times \\ &\left[\frac{48\pi^{3/2}MS^5}{e^{\left(\frac{-q^2 \sqrt{\pi}}{2M \sqrt{S}}\right)} \left( 32MPS^{9/2} e^{\left(\frac{q^2 \sqrt{\pi}}{2M \sqrt{S}}\right)} -24\pi^{3/2} M^2S^3 + 24\pi^2Mq^2S^{3/2} - 3\pi^{5/2}q^4S^2\right)} \right]  \end{split}    
\end{equation}
\begin{figure}
    \centering
    \includegraphics[width=10cm,height=7cm]{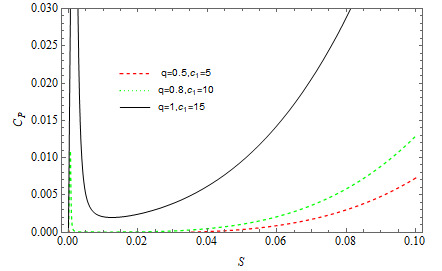}
    \caption{$C_P$ in terms of entropy \emph{S}. Here we have taken \emph{m} = 0.2.}
    \label{fig:3}
\end{figure}
The phase structure expected in such a scenario with charged black holes in massive gravity is that of a Van der Waal fluid system \cite{48}. We can see that the system undergoes a phase transition for smaller black holes and the positive values of $C_P$ for smaller as well as larger black holes is suggestive of their stability.

\section{Efficiency of the RCB heat engine}
\label{section :3}
\begin{figure}
    \centering
{{\includegraphics[width=7cm,height=5cm]{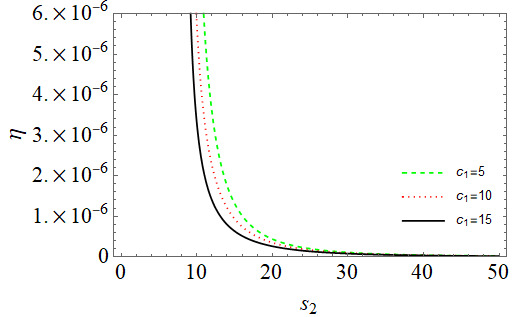} }}
\qquad
{{\includegraphics[width=7cm,height=5cm]{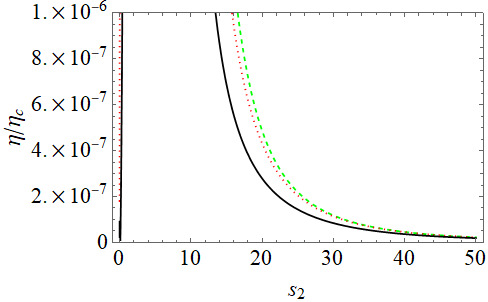} }}
\caption{Efficiency of RCB heat engine with varying entropy \emph{S}. The black solid line represents $c_1 = 15$ and the green and red dotted lines represents $c_1 = 5$ and $c_1 = 10$ respectively. The graviton mass is taken to be m = 0.2 and other parameters are taken to be q = 1, $P_1 = 4$, $P_4$ = 1, $S_2$ = 12, $S_1$ = 8.  }
\label{fig:4}
\end{figure}

The work done during the entire cycle can be obtained by the usual corresponding relation from thermodynamics,
\begin{equation*}
    W = \oint PdV
\end{equation*} which can be written in the form of \emph{P} and \emph{S} using the expressions $S = \pi r_+^2$ \cite{2} and $V = \frac{4\pi r_+^2}{3}$,
\begin{equation} \label{eq : 10}
    W_{total} = \frac{4}{3\sqrt{\pi}} (P_1 - P_4) \left(S_2^{3/2} - S_1^{3/2}\right) 
\end{equation} The net input heat can be given as,
\begin{equation*}
    Q_H = \int_{T_1}^{T_2} C_P dT
\end{equation*} which can be written in terms of \emph{T} and \emph{S} as,
\begin{equation}
\begin{split} \label{eq : 11}
    Q_H &= \frac{\pi}{6P_1^2}(T_2^3 - T_1^3) = \frac{4}{3\sqrt{\pi}}P_1 (S_2^{3/2} -  S_1^{3/2}) \\ &= \biggl\{\ \biggl[32\sqrt{\pi}P_1 e^{\left(\frac{\sqrt{\pi}q^2}{2M \sqrt{S_1}}\right)}S_2^{3/2} + 9\pi c_1^2 me^{\left(\frac{\sqrt{\pi}q^2}{2M \sqrt{S_1}}\right)}S_2 + \biggl(-32\sqrt{\pi}P_1S_1^{3/2} - 9\pi c_1^2mS_1 + \\ & \left(36\pi^2 \Gamma\left(0 , \sqrt{\pi}q^2/2M S_2\right) - 36\pi^2 \Gamma\left(0 , \sqrt{\pi}q^2/2M S_1\right)\right) \biggr)e^{\left(\frac{\sqrt{\pi}q^2}{2M \sqrt{S_1}}\right)} + 36\pi^2 M \biggr]e^{\left(\frac{\sqrt{\pi}q^2}{2M \sqrt{S_2}}\right)} \\ &- 36\pi^2Me^{\left(\frac{\sqrt{\pi}q^2}{2M \sqrt{S_1}}\right)} \biggr\} e^{\left(\frac{-\sqrt{\pi}q^2}{2M \sqrt{S_2}} - \frac{\sqrt{\pi}q^2}{2M \sqrt{S_1}} \right)}
\end{split}
\end{equation} The efficiency of the cycle can be written using \ref{eq : 10} and \ref{eq : 11} as, 
\begin{equation}
\begin{split}
\eta &= \frac{W}{Q_H} = 1-\frac{P_4}{P_1} = 1 - \frac{T_C}{T_H} \\  &= \left( 1 - \frac{P_4}{P_1}\right)\times P_1 \ (S_2^{3/2} -  S_1^{3/2}) \biggl\{\frac{3\sqrt{\pi}}{4} \biggl[32\sqrt{\pi}P_1 e^{\left(\frac{\sqrt{\pi}q^2}{2M \sqrt{S_1}}\right)}S_2^{3/2} + 9\pi c_1^2 me^{\left(\frac{\sqrt{\pi}q^2}{2M \sqrt{S_1}}\right)}S_2 + \\   \ &\ \ \biggl(-32\sqrt{\pi}P_1S_1^{3/2} - 9\pi c_1^2mS_1 +  \left(36\pi^2 \Gamma\left(0 , \sqrt{\pi}q^2/2M S_2\right) - 36\pi^2 \Gamma\left(0 , \sqrt{\pi}q^2/2M S_1\right)\right)\biggr) \\ &\times \  \ e^{\left(\frac{\sqrt{\pi}q^2}{2M \sqrt{S_1}}\right)} + 36\pi^2 M \biggr]e^{\left(\frac{\sqrt{\pi}q^2}{2M \sqrt{S_2}}\right)} - \\ &\   \ 36\pi^2Me^{\left(\frac{\sqrt{\pi}q^2}{2M \sqrt{S_1}}\right)}\biggr\}^{-1} \left[e^{\left(\frac{-\sqrt{\pi}q^2}{2M \sqrt{S_2}} - \frac{\sqrt{\pi}q^2}{2M \sqrt{S_1}} \right)}\right]^{-1} 
\end{split}
\end{equation} 
It would be better if we can compare this efficiency with the maximum efficiency obtainable for a heat engine, i.e., the Carnot efficiency. Its expression is obtained as follows,
\begin{equation}
\begin{split}
\eta_c &= 1 - \frac{T_4 (S_1,P_4)}{T_2 (S_2,P_1)} \\ &= 1 - \frac{\frac{1}{4\sqrt{\pi S_1}}\left[ \frac{2M\sqrt{\pi}}{\sqrt{S_1}}e^{\left( \frac{-q^2\sqrt{\pi}}{2M\sqrt{S_1}}\right)} - \frac{q^2\pi}{S_1}e^{\left( \frac{-q^2\sqrt{\pi}}{2M\sqrt{S_1}}\right)} + \frac{16S_1P_4}{3\pi} +c^2m\sqrt{\frac{S_1}{\pi}} \right]}{\frac{1}{4\sqrt{\pi S_2}}\left[ \frac{2M\sqrt{\pi}}{\sqrt{S_2}}e^{\left( \frac{-q^2\sqrt{\pi}}{2M\sqrt{S_2}}\right)} - \frac{q^2\pi}{S_2}e^{\left( \frac{-q^2\sqrt{\pi}}{2M\sqrt{S_2}}\right)} + \frac{16S_2P_1}{3\pi} +c^2m\sqrt{\frac{S_2}{\pi}} \right]}
\end{split}
\end{equation}
With these expressions at hand, we next turn to a detailed analysis of the efficiency of the system under consideration which is the main motive of the paper. The plots of efficiency under varying graviton mass \emph{m} in addition to the changing entropy \emph{S} and pressure \emph{P} helps us fully understand the influence of massive gravity parameters on the efficacy of a RCB heat engine. From Figure \ref{fig:4} we see that the efficiency decreases exponentially and reaches a minimum value and after that becomes zero monotonically with increasing entropy. This sharp decrease in efficiency is slightly greater for larger values of $c_1$ parameter. Thus we can say that  increasing the $c_1$ parameter leads to decrease in $\eta$. Similar behaviour is observed for the plot of ${\eta}/{\eta_c}$ on the right side, where there is an exponential decrease in the ratio ${\eta}/{\eta_c}$ and then a monotonic decrease thereafter. Also the ratio decreases much faster for increasing values of $c_1$.      
\par
\begin{figure}
    \centering
\subfloat[]{{\includegraphics[width=7cm,height=5cm]{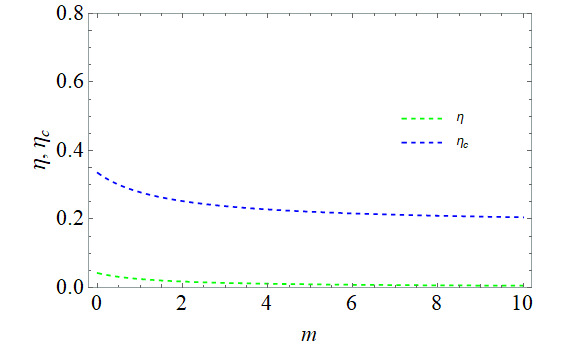} }}
\qquad
\subfloat[]{{\includegraphics[width=7cm,height=5cm]{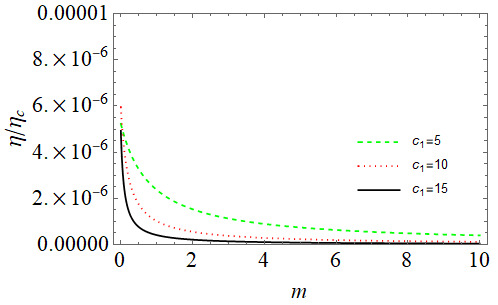} }}
\caption{Efficiency of RCB heat engine with varying graviton mass \emph{m}. a) The green and red dotted line represents RCB efficiency and Carnot efficiency respectively. b) The black solid line represents $c_1 = 15$ and the green and red dotted lines represents $c_1 = 5$ and $c_1 = 10$ respectively. The graviton mass is taken to be m=1. }
\label{fig:6}
\end{figure}

 Figure \ref{fig:5} below shows the plot of efficiency in terms of pressure which implies that there is a monotonous increase in the efficiency with pressure. This could mean that the efficiency would reach to values very closer to its maximum value as it approaches infinity. This can also be seen from the $\eta/ \eta_c$ plot of RCB heat engine efficiency  and the Carnot efficiency where these values never really become equal.
\begin{figure}
    \centering
{{\includegraphics[width=7cm,height=5cm]{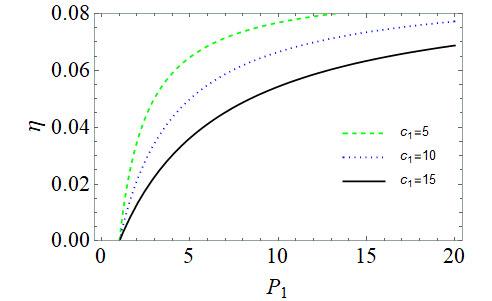} }}
\qquad
{{\includegraphics[width=7cm,height=5cm]{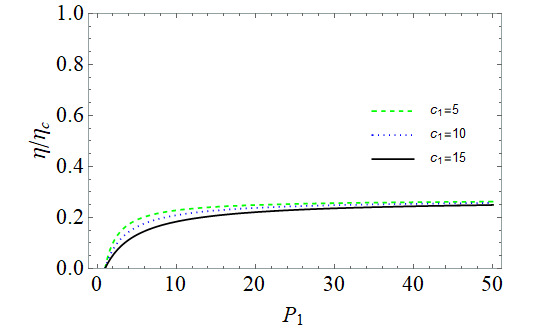} }}
\caption{Efficiency of RCB heat engine in massive gravity with varying pressure $P_1$. The graviton mass is taken to be m = 0.2, q = 1, $P_1 = 4$, $P_4$ = 1, $S_2$ = 12, $S_1$ = 8. }
\label{fig:5}
\end{figure}Now we come to the interesting part where we observe the effect of massive graviton on the RCB heat engine efficiency by plotting the same against increasing value of graviton mass \emph{m}. From figure \ref{fig:6} it is evident that the efficiency of the heat engine keeps on decreasing monotonically with increasing value of \emph{m}. We can also see that the maximum efficiency one can obtain in such a system is around 0.4 which can be considered too low. But when compared to the efficiency of the heat engine without the massive gravity parameter, we see that graviton mass definitely increases the efficacy of the system considered. 
  
\newpage
\section{Conclusion}
It has been shown before that charged black holes can be studied as a thermodynamic heat engine when considered under the context of a modified version of gravity, i.e., massive gravity. The interesting result from such a study is that the graviton mass actually increases the efficiency of the RCB heat engine. We study the influence of addition of a mass to the graviton on the efficiency of heat engine in case of regular charged black holes. Firstly we have constructed  RCB heat engine within the framework of massive gravity by adding massive graviton terms to the metric function. There are two cases that we have considered, one where the value of graviton mass is fixed to be \emph{m} = 0.2  and other case where the graviton mass is kept changing. In the first case, we start with examining the critical behaviour of the system with the help of a very important thermodynamic quantity, the specific heat capacity. It can be seen that quantities like Hawking's temperature, pressure and heat capacity are functions of graviton mass. We then move on to calculate the efficiency of RCB heat engine and study in detail how it varies with quantitites like entropy and pressure through the plots of efficiency against varying parameters of \emph{q} and \emph{c} with fixed graviton mass term. The graphical analysis suggests that the efficiency is a decreasing function of entropy and increasing function of pressure. Also the value of the efficiency is quite low. We further analyse the  efficiency by varying graviton mass \emph{m} to highlight the influence of massive gravity and find out that the mass term increases the efficiency of the RCB heat engine.    


\begin{thebibliography}{0}
\bibitem{1}Israel, Werner. "Event horizons in static vacuum space-times." Physical review 164.5 (1967): 1776.
\bibitem{2}Bekenstein, Jacob D. "Black holes and the second law." JACOB BEKENSTEIN: The Conservative Revolutionary. 2020. 303-306.
\bibitem{3}Bekenstein, Jacob D. "Generalized second law of thermodynamics in black-hole physics." Physical Review D 9.12 (1974): 3292.
\bibitem{4}Hawking, Stephen W. "Particle creation by black holes." Euclidean quantum gravity. 1975. 167-188.
\bibitem{5}Bardeen, James M., Brandon Carter, and Stephen W. Hawking. "The four laws of black hole mechanics." Communications in mathematical physics 31.2 (1973): 161-170.
\bibitem{6}Jacobson, Ted. "Thermodynamics of spacetime: the Einstein equation of state." Physical Review Letters 75.7 (1995): 1260.
\bibitem{7}Padmanabhan, Thanu. "Thermodynamical aspects of gravity: new insights." Reports on Progress in Physics 73.4 (2010): 046901.
\bibitem{8}Hawking, Stephen W., and Don N. Page. "Thermodynamics of black holes in anti-de Sitter space." Communications in Mathematical Physics 87.4 (1983): 577-588.
\bibitem{9}Maldacena, Juan. "The large-N limit of superconformal field theories and supergravity." International journal of theoretical physics 38.4 (1999): 1113-1133.
\bibitem{10}Witten, Edward. "Anti-de Sitter space and holography." Advances in Theoretical and Mathematical Physics 2 (1998): 253-291.
\bibitem{11}Witten, Edward. "Anti-de Sitter space, thermal phase transition, and confinement in gauge theories." THE OSKAR KLEIN MEMORIAL LECTURES 1988–1999. 2014. 389-419.
\bibitem{12}Kovtun, Pavel K., Dan T. Son, and Andrei O. Starinets. "Viscosity in strongly interacting quantum field theories from black hole physics." Physical review letters 94.11 (2005): 111601.
\bibitem{13}Hartnoll, Sean A., et al. "Theory of the Nernst effect near quantum phase transitions in condensed matter and in dyonic black holes." Physical Review B 76.14 (2007): 144502.
\bibitem{14}Hartnoll, Sean A., Christopher P. Herzog, and Gary T. Horowitz. "Building a holographic superconductor." Physical Review Letters 101.3 (2008): 031601.
\bibitem{15}Gruber, Christine, Orlando Luongo, and Hernando Quevedo. "Geometric approaches to the thermodynamics of black holes." The Fourteenth Marcel Grossmann Meeting On Recent Developments in Theoretical and Experimental General Relativity, Astrophysics, and Relativistic Field Theories: Proceedings of the MG14 Meeting on General Relativity, University of Rome “La Sapienza”, Italy, 12–18 July 2015. 2018.
\bibitem{16}Teitelboim, Claudio. "The cosmological constant as a thermodynamic black hole parameter." Physics Letters B 158.4 (1985): 293-297.
\bibitem{17}Brown, J. David, and Claudio Teitelboim. "Neutralization of the cosmological constant by membrane creation." Nuclear Physics B 297.4 (1988): 787-836.
\bibitem{18}Creighton, Jolien DE, and Robert B. Mann. "Quasilocal thermodynamics of dilaton gravity coupled to gauge fields." Physical Review D 52.8 (1995): 4569.
\bibitem{19}Dolan, Brian P. "The cosmological constant and the black hole equation of state." Classical and Quantum Gravity 28.12 (2010): 125020.
\bibitem{20}Cvetič, Mirjam, et al. "Black hole enthalpy and an entropy inequality for the thermodynamic volume." Physical Review D 84.2 (2011): 024037.
\bibitem{21}Kastor, David, Sourya Ray, and Jennie Traschen. "Enthalpy and the mechanics of AdS black holes." Classical and Quantum Gravity 26.19 (2009): 195011.
\bibitem{22}Chamblin, Andrew, et al. "Charged AdS black holes and catastrophic holography." Physical Review D 60.6 (1999): 064018.
\bibitem{23}Chamblin, Andrew, et al. "Holography, thermodynamics, and fluctuations of charged AdS black holes." Physical Review D 60.10 (1999): 104026.
\bibitem{24}Gunasekaran, Sharmila, David Kubizňák, and Robert B. Mann. "Extended phase space thermodynamics for charged and rotating black holes and Born-Infeld vacuum polarization." Journal of High Energy Physics 2012.11 (2012): 1-43.
\bibitem{25}Frassino, Antonia M., et al. "Multiple reentrant phase transitions and triple points in Lovelock thermodynamics." Journal of High Energy Physics 2014.9 (2014): 1-47.
\bibitem{26}Hennigar, Robie A., and Robert B. Mann. "Reentrant phase transitions and van der Waals behaviour for hairy black holes." Entropy 17.12 (2015): 8056-8072.
\bibitem{27}Altamirano, Natacha, et al. "Kerr-AdS analogue of triple point and solid/liquid/gas phase transition." Classical and Quantum Gravity 31.4 (2014): 042001.
\bibitem{28}Wei, Shao-Wen, and Yu-Xiao Liu. "Triple points and phase diagrams in the extended phase space of charged Gauss-Bonnet black holes in AdS space." Physical Review D 90.4 (2014): 044057.
\bibitem{29}Johnson, Clifford V. "Holographic heat engines." Classical and Quantum Gravity 31.20 (2014): 205002.
\bibitem{30}Fierz, Markus, and Wolfgang Ernst Pauli. "On relativistic wave equations for particles of arbitrary spin in an electromagnetic field." Proceedings of the Royal Society of London. Series A. Mathematical and Physical Sciences 173.953 (1939): 211-232.
\bibitem{31}van Dam, Hendrik, and Martinus Veltman. "Massive and mass-less Yang-Mills and gravitational fields." Nuclear Physics B 22.2 (1970): 397-411.
\bibitem{32}Zakharov, Valentin I. "LINEARIZED GRAVITATION THEORY AND THE GRAVITON MASS." JETP Lett.(USSR)(Engl. Transl.) 12: 312-14 (5 Nov 1970). (1970).
\bibitem{33}Vainshtein, Arkady I. "To the problem of nonvanishing gravitation mass." Physics Letters B 39.3 (1972): 393-394.
\bibitem{34}Babichev, E., C. Deffayet, and R. Ziour. "Recovering General Relativity from massive gravity." Physical review letters 103.20 (2009): 201102.
\bibitem{35}Niedermann, Florian, and Antonio Padilla. "Gravitational mechanisms to self-tune the cosmological constant: obstructions and ways forward." Physical review letters 119.25 (2017): 251306.
\bibitem{36}Chkareuli, Giga, and David Pirtskhalava. "Vainshtein mechanism in Λ3-theories." Physics Letters B 713.2 (2012): 99-103.
\bibitem{37}Katsuragawa, Taishi, et al. "Relativistic stars in de Rham-Gabadadze-Tolley massive gravity." Physical Review D 93.12 (2016): 124013.
\bibitem{38}Saridakis, Emmanuel N. "Phantom crossing and quintessence limit in extended nonlinear massive gravity." Classical and Quantum Gravity 30.7 (2013): 075003.
\bibitem{39}Cai, Yi-Fu, Caixia Gao, and Emmanuel N. Saridakis. "Bounce and cyclic cosmology in extended nonlinear massive gravity." Journal of Cosmology and Astroparticle Physics 2012.10 (2012): 048.
\bibitem{40}Vegh, David. Holography without translational symmetry. No. arXiv: 1301.0537. 2013.
\bibitem{41}Gümrükçüoğlu, A. Emir, Chunshan Lin, and Shinji Mukohyama. "Open FRW universes and self-acceleration from nonlinear massive gravity." Journal of Cosmology and Astroparticle Physics 2011.11 (2011): 030.
\bibitem{42}Kobayashi, Tsutomu, et al. "New cosmological solutions in massive gravity." Physical Review D 86.6 (2012): 061505.
\bibitem{43}Will, Clifford M. "The confrontation between general relativity and experiment." Living reviews in relativity 17.1 (2014): 1-117.
\bibitem{44}Bhamidipati, Chandrasekhar, and Pavan Kumar Yerra. "Heat engines for dilatonic Born–Infeld black holes." The European Physical Journal C 77.8 (2017): 1-15.
\bibitem{45}Yerra, Pavan Kumar, and Chandrasekhar Bhamidipati. "Heat engines at criticality for nonlinearly charged black holes." Modern Physics Letters A 34.27 (2019): 1950216.
\bibitem{46}Yerra, Pavan Kumar, and Chandrasekhar Bhamidipati. "Critical heat engines in massive gravity." Classical and Quantum Gravity 37.20 (2020): 205020.
\bibitem{47}Culetu, Hristu. "On a regular modified Schwarzschild spacetime." arXiv e-prints (2013): arXiv-1305.
\bibitem{48}Hendi, S. H., et al. "Van der Waals like behavior of topological AdS black holes in massive gravity." Physical Review D 95.2 (2017): 021501.
\end{thebibliography}
\end{document}